\documentclass[preprint2]{aastex}

\usepackage{graphics}

\newcommand{\etal}{{\it et al.}\ }

\slugcomment{To appear in the {\it Pubblications of the Astronomical Society of the Pacific}}
\shorttitle{Lick spectral indices for super metal-ricj stars}
\shortauthors{A. Buzzoni, M. Chavez, M.L. Malagnini, and C. Morossi}

\begin{document}

\title{Lick Spectral Indices for Super Metal-rich Stars\footnote{Based on observations 
collected at the INAOE ``G. Haro'' Observatory, Cananea (Mexico)}}

\author{A. Buzzoni}
\affil{Telescopio Nazionale Galileo, A.P. 565, 38700 Santa Cruz de La Palma, Canary Islands, Spain,\\
and Osservatorio Astronomico di Brera, Milano Italy}
\email{buzzoni@tng.iac.es}

\author{M. Chavez}
\affil{Instituto Nacional de Astrof\'\i sica, Optica y Electr\'onica, A.P. 51 y 216, 72000 Puebla, Mexico}
\email{mchavez@inaoep.mx}

\author{M.L. Malagnini}
\affil{Dipartimento di Astronomia, Universit\`a di Trieste, Via G.B. Tiepolo 11, 34131 Trieste, Italy}
\email{malagnini@ts.astro.it}

\author{C. Morossi}
\affil{Osservatorio Astronomico di Trieste, Via G.B. Tiepolo 11, 34131 Trieste, Italy}
\email{morossi@ts.astro.it}

\begin{abstract}
We present Lick spectral indices for a complete sample of
139 candidate super metal-rich stars of different luminosity class
(MK type from V to I). For 91 of these stars we were
able to identify, in an accompanying paper (Malagnini \etal 2000),
the fundamental atmosphere parameters.
This confirms that at least 2/3 of the sample consists of stars with [Fe/H] in 
excess of +0.1~dex.

Optical indices for both observations and fiducial synthetic spectra
have been calibrated to the Lick system according to Worthey \etal (1994), 
and include the \ion{Fe}{1} indices of Fe5015, Fe5270, and Fe5335, and the \ion{Mg}{1} and MgH indices
of Mg$_2$ and Mgb at 5180 \AA.

Internal accuracy of the observations is found to be 
$\sigma{\rm (Fe5015)} = \pm 0.32$~\AA, $\sigma{\rm (Fe5270)} = \pm 0.19$~\AA, 
$\sigma{\rm (Fe5335)} = \pm 0.22$~\AA, $\sigma$(Mg$_2$)~$= \pm 0.004$~mag, and
$\sigma$(Mgb)~$= \pm 0.19$~\AA. This is about a factor of two better
than the corresponding theoretical indices from the synthetic spectra, 
the latter as a consequence of the intrinsic limitations in the input physics,
as discussed in Chavez \etal (1997).

By comparing models and observations, no evidence is found for non-standard
Mg vs.\ Fe relative abundance, so that [Mg/Fe] = 0, on the average, 
for our sample.

Both the Worthey \etal (1994) and Buzzoni \etal (1992, 1994) fitting functions
are found to suitably match the data, and can therefore confidently be extended 
for population synthesis application also to super-solar metallicity regimes.
A somewhat different behaviour of the two fitting sets appears, however, beyond the 
temperature constraints of our stellar sample. Its impact on the theoretical
output is discussed, as far as the integrated Mg$_2$ index is derived from
synthesis models of stellar aggregates.

A two-index plot, like Mg$_2$ vs.\ Fe5270, is found to provide a simple and
powerful tool to probe distinctive properties of single stars and
stellar aggregates as a whole. The major advantage, over a classical c-m 
diagram, is to be both reddening free and distance independent.

\end{abstract}


\section{Introduction}

A fair recognition of super metal-rich (SMR) stars
in the solar neighborhood remains a crucial issue to confidently assess the 
problem of metallicity estimate in the Galaxy bulge and external
galaxies, via population synthesis models.

In the classical works on the subject (Spinrad and Taylor 1969), the 
definition of SMR candidates mostly relied on the relative chemical abundance 
with respect to the Hyades and set the nominal threshold for SMR stars at 
[Fe/H]~$> +0.2$~dex (that is a metal abundance $Z$ in excess of 50\% 
on the solar abundance).

Both the metallicity scale for the bulge stellar population 
(Frogel 1999) as well as the individual estimates of $Z$ for the most 
metal-rich template stars in the disk (first of all the classical SMR 
standard $\mu$ Leo, cf.\ McWilliam 1997; Taylor 1999; Smith and Ruck 2000)
have however been the subject of important revisions in the recent literature,
although no conclusive argument seems so far to definitely settle the problem.

In this paper, we follow on the spectroscopic study of a sample of SMR candidates
undertaken in Malagnini \etal (2000, hereafter Paper I) providing
a detailed calibration for the main Lick narrow-band 
indices in the optical wavelength range. These include the two Mg indices 
(Mg$_2$ and Mgb) and the three \ion{Fe}{1} indices of Fe5015, Fe5270, and Fe5335
(hereafter referred to in their abbreviated form, i.e.\ Fe50, Fe52
and Fe53, respectively; cf.\ Sec.\ 2).

For most of these stars, we identified in Paper I the photospheric
{\it fiducial} parameters ($T_{\it eff},\ \log g$ and [M/H]) and computed
the corresponding synthetic spectra according to the Chavez \etal (1997)
theoretical database. Theoretical narrow-band indices will be especially dealt
with in Sec.\ 3, where we address the problem of a possible
non-solar [Mg/Fe] chemical partition of our SMR candidates.

Metallicity estimate in external galaxies should forcedly rely on 
population synthesis models by matching integrated indices for the
whole stellar aggregate. This is usually done by means of the so-called
``fitting function'' technique (Worthey \etal 1994, hereafter W94; 
Buzzoni \etal 1992, B92, and Buzzoni \etal 1994, B94), that is by fitting 
index behaviour for local stars and implementing this analytical framework in 
the synthesis code. It is of paramount importance, in this sense, to 
confidently validate fitting function predictions also to super-solar
metallicity regimes. This is done in Sec.\ 4, where we check our
observations against the W94, B92 and B94 fitting sets.

The Mg$_2$ vs.\ Fe52 plot could reveal a simple and powerful tool for
stellar diagnostics both for individual objects and stellar aggregates.
Its major advantage over the classical c-m diagrams is to be both
reddening free and distance independent. In Sec.\ 5 we will discuss
in some detail its possible applications by referring to our
stellar sample.

\section{Observations and system calibration}

The original set of observations of this work consists of
139 bright stars ($V < 8$ mag) in the spectral range between F and M and 
luminosity class between I and V, selected from three main reference sources 
for metallicity in the literature. 

Dwarf stars (class V-IV) mainly come from the general catalog of
Cayrel de Strobel \etal (1997), while giants (class III to I) are mostly
from Taylor (1991). This two data sources have then be completed with the work 
of W94 in order to include a set of primary calibrators
to the Lick system, and provide a more homogeneous distribution among all 
luminosity classes. 

Each of our SMR candidates is quoted to have {\it at least} one determination 
of [Fe/H]~$\geq +0.1$ from high-resolution abundance studies in the literature.

This working sample has been observed during three runs, between December 1995 
and August 1996, at the 2.12m f/12 telescope of the INAOE ``G. Haro'' 
Observatory in Cananea (Mexico). We collected B\"oller and Chivens 
mid--dispersion (35\,\AA/mm) spectra at $R = \lambda/\Delta \lambda = 2000$ 
inverse resolution (namely some 2.5\,\AA\ FWHM) in the 
wavelength range between 4600 and 5500 \AA. Full details of the observing 
setup as well as of the data reduction can be found in Paper I.

A total of 100 stars had two or more observations along different
nights while  39 stars were observed only once. 
Typically, a pixel-to-pixel S/N ratio greater than 50 was achieved 
in the 250 observed spectra.

Wavelength calibration eventually provided a $\pm 0.09$\,\AA\ accuracy, 
while spectral energy distribution (SED) of stars was obtained within a 
$\pm 10\%$ relative flux level. Flux rebinning to a fixed $\Delta \lambda$/pixel
ratio was performed preserving the total number of pixels as in the original 
observations.

Multiple observations also allowed us an independent estimate of the internal 
uncertainty  of the whole data set. They confirmed in particular a
pixel-to-pixel photon noise better than $\pm 0.02$\ mag, in agreement with the
observed S/N ratio, with the only relevant exception of star
HD 36389, which had poorer observations and S/N = 6 (cf. Table 1 in Paper I).
Repeated observations for each star were eventually co-added
after calibration in order to obtain one mean spectrum for each object.

Table 1 reports a full summary of the observed stellar sample listing in 
columns (1)--(4) respectively the HD number, spectral type, number of collected 
spectra, and mean flux uncertainty as from the standard deviation of the
relative flux for repeated observations.\footnote{For single observations, 
a mean $\sigma(\rm flux) = 0.02$ will be assumed in this work.}

\subsection {Matching the Lick system}

Three striking features of \ion{Fe}{1} and the atomic-molecular blend
of \ion{Mg}{1} and MgH are present in the wavelength range of our spectra.

Each of these features has been included in the narrow-band spectral
indices of the Lick system (Faber \etal 1977), in its recent extension
by W94. A measure of the line strength is provided in 
this system by interpolating a local continuum ($f_c$) from two side bands 
adjacent to each relevant feature, and evaluating therefrom a pseudo-equivalent
width ($E.W.$) such as
\begin{equation}
\begin{array}{lll}
E.W. & = & \int [(f_c(\lambda) - f(\lambda))/ f_c(\lambda)] d\lambda \\
     & = & \Delta\lambda_f\ \left( 1 - < {f / f_c} > \right)
\end{array}
\end{equation}
where $\Delta\lambda_f$ is the width of the spectral window centered on the
absorption feature.

For some molecular blends, it is useful to define an index in magnitude 
scale ($I_{mag}$) so that
\begin{equation}
\begin{array}{lll}
I_{mag} & = & -2.5\ \log \left( \int [f(\lambda) / f_c(\lambda)] d\lambda\ /\Delta\lambda_f \right) \\
        & = & -2.5\ \log < f / f_c >
\end{array}
\label{eq:index}
\end{equation}

Both formalisms are however equivalent (cf. also Brodie and Huchra 1990) as
\begin{equation}
E.W. = \Delta \lambda_f\ \left(1 - 10^{-0.4\ I_{mag}}\right).
\end{equation}

For each star in our sample, the Lick indices of Fe50, Fe52, Fe53,
Mg$_2$ and Mgb have been computed in the instrumental system, following W94.

To account for the different spectral resolution of our observations,
calibration to the standard system has been accomplished by matching
the index calibrators in common with W94, B92 and B94, using the latter
ones as a ``check sample'' in our fit.

We especially restrained our analysis to stars of class V-IV-III since
class I-II supergiants are often variable stars and are therefore 
less confident calibrators.

A total of 36 stars were found in common with W94, with all
available indices,\footnote{Apart from HD 49161, that lacks Fe50
in the W94 sample.}
while 13 secondary calibrators appeared in the B92 and B94 sample with 
standard Mg$_2$, Fe52 and Fe53 index measurements.

A zero-average (O--C) residual distribution (in the sense 
[Observed -- Standard]) was eventually obtained
with the following set of transformation equations:
\begin{equation}
\begin{array}{lc}
(O-C)_{Fe50} = 0.139\ {\rm Fe50} -0.278 & [{\rm \AA}] \\
(O-C)_{Mg2} =  -0.056\ {\rm Mg}_2 +0.007 & [{\rm mag}] \\
(O-C)_{Mgb} = 0.047\ {\rm Mg}_b -0.092 & [{\rm \AA}] \\
(O-C)_{Fe52} = 0.150\ {\rm Fe52} -0.183 & [{\rm \AA}] \\
(O-C)_{Fe53} = 0.163\ {\rm Fe53} +0.078 & [{\rm \AA}]
\end{array}
\label{eq:oc}
\end{equation}

In the equations, the input indices are the observed ones (that is in
the instrumental system). After calibration, the residual standard 
deviation for the W94 primary calibrators and the B92/B94 ``check sample'', resulted
\begin{equation}
\begin{array}{ll}
\sigma({\it Fe50}) = \pm 0.39\ {\rm \AA} & {\rm 35\ stars} \\
\sigma({\it Mg}_2) = \pm 0.012\ {\rm mag} & {\rm 49\ stars} \\
\sigma({\it Mg}b) = \pm 0.19\ {\rm \AA} & {\rm 36\ stars} \\
\sigma({\it Fe52}) = \pm 0.24\ {\rm \AA} & {\rm 49\ stars} \\
\sigma({\it Fe53}) = \pm 0.25\ {\rm \AA} & {\rm 49\ stars.}
\end{array}
\label{eq:calib}
\end{equation}

\begin{figure}
\resizebox{\hsize}{!}{\includegraphics{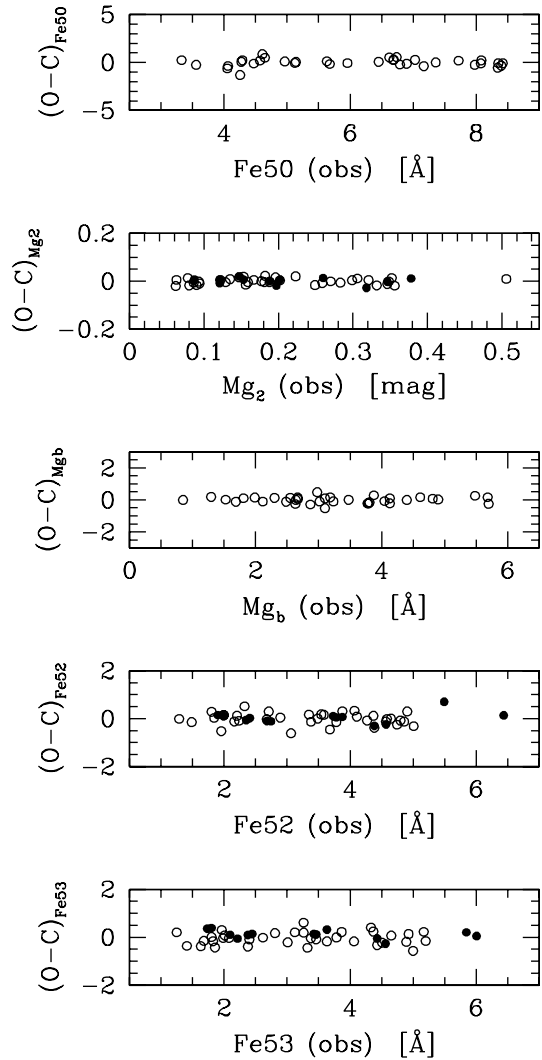}}
\caption{Index calibration to the standard Lick system. Residuals for
the 36 primary calibrators (of luminosity class V-IV-III) in common with 
W94 ($\circ$) and a ``check sample'' of 13 stars in common with B92 and B94
($\bullet$) are displayed, after application of eq.~(\ref{eq:oc}). Point 
scatter is $\sigma({\it Fe50}) = \pm 0.39\ {\rm \AA},\ \sigma({\it Mg}_2) 
= \pm 0.012\ {\rm mag},\ \sigma({\it Mg}b) = \pm 0.19\ {\rm \AA},\ 
\sigma({\it Fe52}) = \pm 0.24\ {\rm \AA}$, and $\sigma({\it Fe53}) 
= \pm 0.25\ {\rm \AA}$. These values are comparable with the internal
uncertainty in the observations confirming the reliability of the
calibration procedure.}
\label{oc}
\end{figure}

Figure~\ref{oc} summarizes our results while the complete list of
standard indices is reported in columns (5)--(9) of Table 1.
When available from Paper I, column (10) provides the fiducial metallicity,
while column (11) marks the reference calibrators in common with W94, 
B92 and B94.

A convenient estimate of the internal uncertainty of the whole data sample
can easily derive from the typical pixel-to-pixel S/N ratio, recalling that 
$\sigma({\rm flux}) = (S/N)^{-1}$~mag.
For each index, therefore, in a magnitude scale we could write
\begin{equation}
\sigma_I = {{\sigma({\rm flux)}}\over {\sqrt{N_{red}+N_{blue}} + \sqrt {N_{feat}}}} \quad {\rm [mag]}
\end{equation}
where $N_{red}$, $N_{blue}$ and $N_{feat}$ are the number
of pixels in the red and blue continuum side bands, and in the feature
window, respectively.
To translate from magnitudes to \AA\ equivalent width, we simply have
\begin{equation}
\sigma\ {\rm [\AA]} = \sigma\ {\rm [mag]} \times \Delta\lambda_f.
\end{equation}

The typical internal error bar of our data eventually resulted in
\begin{equation}
\begin{array}{l}
\sigma({\it Fe50}) = \pm 0.32\ {\rm \AA} \\
\sigma({\it Mg}_2) = \pm 0.004\ {\rm mag} \\
\sigma({\it Mg}b) = \pm 0.19\ {\rm \AA}  \\
\sigma({\it Fe52}) = \pm 0.19\ {\rm \AA} \\
\sigma({\it Fe53}) = \pm 0.22\ {\rm \AA.}
\end{array}
\end{equation}

Note that the latter quantities consistently compare with the 
point scatter of eq.~(\ref{eq:calib}). This confirms that no residual variance 
was left in the data after transformation to the standard system 
via eq.~(\ref{eq:oc}).

\section{The fiducial synthetic spectra}

For 73 of the stars in Table 1 (nearly all those of luminosity
classes V--III), we identified in Paper I a best choice for the
atmosphere fundamental parameters available from high-resolution studies 
in the literature. In the following, this will be referred to as a
``fair'' sample. A further subsample of 18 stars had a formal atmosphere
solution in our work, but with lower statistical confidence.
The total of 91 stars will compose our ``extended'' sample.

Operationally, in Paper I our strategy was to compute, for each 
high-resolution parameter set, the corresponding synthetic spectrum from
the Chavez \etal (1997) database and 
pick up the better fit to our mid-resolution spectra. 
This procedure allowed us to assess physical self-consistency of the 
high-resolution outputs and provide the best values for 
$\log T_{\it eff},\ \log g$ and [Fe/H].

In our analysis we had to face of course some intervening limitations
of the Chavez \etal (1997) database, based on  Kurucz (1993) models,
at lower temperature and gravity.
While, from one hand, the plane-parallel layer approximation is no 
longer valid in the model atmosphere for low-gravity stars ($\log g < 1.5$~dex),
the incomplete sampling of the molecular contribution leads, on the other
hand, to a poor representation of the SED in case of stars cooler than 4000~K.

In addition, when dealing with mid-resolution spectra, a study of our
synthetic spectra showed that some unavoidable degeneracy in the allowed range
of the atmosphere physical parameters is present, and
statistically equivalent best fits to the observations may be provided by 
theoretical spectra with slightly different combinations of the distinctive 
parameters.

For example, we verified that the effect of any change in the effective 
temperature of a star can be compensated by a combined change of
gravity and metallicity such as
\begin{equation}
\Delta \log g /\Delta T_{\it eff} = 1.3\ (1000/T_{\it eff})^4~{\rm dex\ K}^{-1},
\end{equation}
and
\begin{equation}
\Delta {\rm [Fe/H]} /\Delta T_{\it eff} = 7 \ 10^{-4}~{\rm dex\ K}^{-1}.
\label{eq:e10}
\end{equation}

Therefore, a warmer fitting temperature would accordingly imply
a higher surface gravity and metallicity.

The [Fe/H] vs.\ $T_{\it eff}$ dependence even more magnifies in case 
we set gravity in the fitting synthetic spectra. Our tests show that
\begin{equation}
\Delta {\rm [Fe/H]} /\Delta T_{\it eff} =  0.2\ (1000/T_{\it eff})^3~{\rm dex\ K}^{-1}. 
\end{equation}

On the other hand, an opposite trend would result between gravity and 
metallicity in case we could fix the fitting temperature:
\begin{equation}
\Delta {\rm [Fe/H]} /\Delta \log g = -0.35
\label{eq:e12}
\end{equation}
that is a higher metal abundance is required to balance ``too shallower''  
absorption features in a spectrum in case of a lower surface gravity.

While therefore our mid-resolution observations alone prevented a univocal
fit to the data, they allowed, though, to discriminate among different
high-resolution parameter sets, and single out the best one for each star.

A total of 64 out of 91 stars in the ``extended'' sample 
resulted to be consistent with a metal abundance ${\rm [Fe/H]} \geq 0.1$~dex
suggesting that about 2/3 of the whole sample of 139 stars are actually 
composed by {\it bona fide} SMR stars.

As for the observed spectra, narrow-band spectral indices have been
computed also for the corresponding set of synthetic spectra,
and transformed to the Lick system by means of the empirical calibration
of eq.~(\ref{eq:oc}).

\begin{figure}
\resizebox{\hsize}{!}{\includegraphics{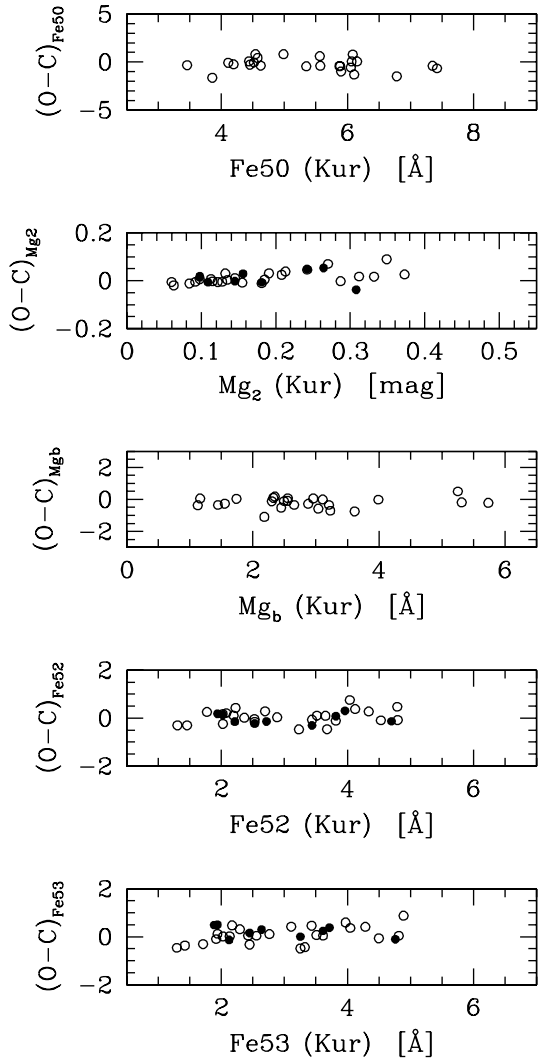}}
\caption{Index calibration to the standard Lick system, after 
eq.~(\ref{eq:oc}), for the fiducial model atmospheres of the 25 primary 
calibrators in common with W94 ($\circ$) and the 9 stars from 
B92 and B94 ($\bullet$). The (O--C) residuals are in the 
sense [Synthetic -- Standard]. Point scatter
is $\sigma({\it Fe50}) = \pm 0.70\ {\rm \AA},\ \sigma({\it Mg}_2) 
= \pm 0.029\ {\rm mag},\ \sigma({\it Mg}b) = \pm 0.40\ {\rm \AA},\ 
\sigma({\it Fe52}) = \pm 0.27\ {\rm \AA}$, and $\sigma({\it Fe53}) 
= \pm 0.34\ {\rm \AA}$. These higher values, compared with 
those of the observations (cf. Fig.~\ref{oc}), give a measure of the 
residual variance of the theoretical models due to the intrinsic limitation 
of the input physics.}
\label{oc_synt}
\end{figure}

In Fig.~\ref{oc_synt} we report the final residuals for the corresponding
subsample of index calibrators with available fiducial model
(the (O--C) residual distribution is now in the sense [Synthetic -- Standard]).
This consists of 25 primary reference stars in the W94 sample and
9 stars from the B92 and B94 list.

Point scatter in Fig.~\ref{oc_synt} is slightly higher compared with 
the observations (especially for the Magnesium indices and Fe50). This can
be regarded as a measure of the ``internal uncertainty'' in the theoretical 
models due to the intrinsic limitation of the input physics:
\begin{equation}
\begin{array}{ll}
\sigma({\it Fe50}) = \pm 0.70\ {\rm \AA} & {\rm 25\ stars} \\
\sigma({\it Mg}_2) = \pm 0.029\ {\rm mag} & {\rm 34\ stars} \\
\sigma({\it Mg}b) = \pm 0.40\ {\rm \AA} & {\rm 25\ stars} \\
\sigma({\it Fe52}) = \pm 0.27\ {\rm \AA} & {\rm 34\ stars} \\
\sigma({\it Fe53}) = \pm 0.34\ {\rm \AA} & {\rm 34\ stars.}
\end{array}
\label{eq:calib_synt}
\end{equation}

\subsection {Index calibration and atmosphere parameters}

A more direct hint of the accuracy of the atmosphere fundamental parameters, 
at least for the 91 stars in our ``extended sample'', can be done by comparing
the (O--C) distribution (in the sense [Observed -- Synthetic]) vs.\ fiducial 
set of temperature, gravity and metallicity.
We especially focussed on the Mg$_2$ index and the ``combined'' Fe index
defined as $<{\rm Fe}> = ({\rm Fe52}+{\rm Fe53})/2$ (Faber \etal 1985).

\begin{figure*}[t]
\resizebox{\hsize}{!}{\includegraphics{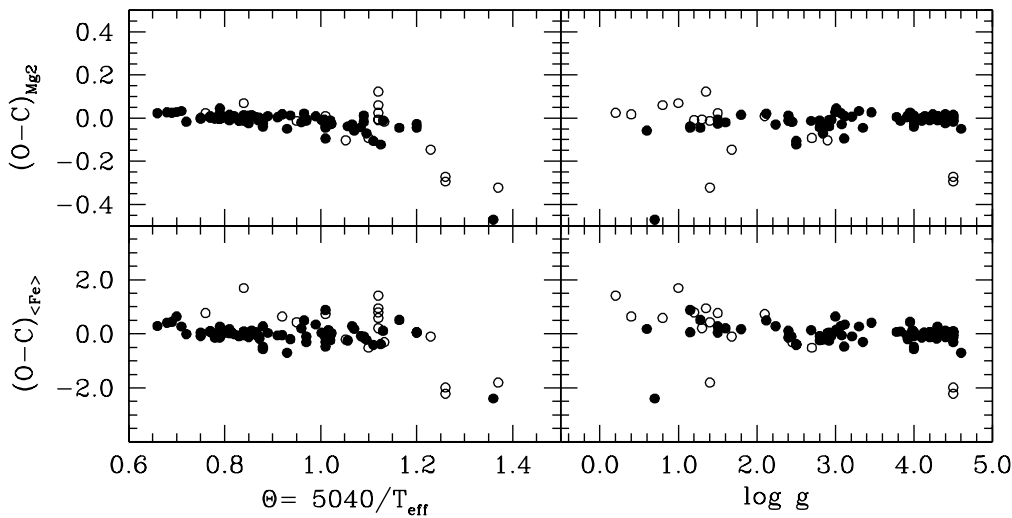}}
\caption{The (O--C) index residuals (in the sense [Observed -- Synthetic])
for Mg$_2$ and combined Iron index $<{\rm Fe}>$ are displayed vs.\ atmosphere 
fiducial parameters $\Theta = 5040/T_{\it eff}$
and surface gravity. Solid dots mark the ``fair sample'' of 73 stars 
while open dots include the supplementary 18 stars with less confident
atmosphere match. As expected, the four coolest stars in the sample 
($T_{\it eff} \leq 4000$~K) are poorly matched by the models while the remaining 
point distribution is consistent with a zero-average residual. These four 
outliers will not enter any further analysis.}
\label{oc_all}
\end{figure*}

The (O--C) trend vs. surface gravity and temperature parameter 
$\Theta = 5040/T_{\it eff}$ is studied in Fig.~\ref{oc_all}.
While a better match is achieved for the ``fair sample'' (solid dots),
an increasing fraction of deviating points appears among cool 
(super)-giant stars (cf. open-dot distribution).

Both in the Mg$_2$ and $<{\rm Fe}>$ plots, four outliers, namely stars 
HD~36389 (sp. type M2Iab), HD~88230 (K8V), HD~146051 (M0.5III), and HD~157881 
(K7V), display a very negative (O--C). These are the four coolest stars in our
sample with $T_{\it eff} \leq  4000$~K. As discussed in 
B92 and Chavez \etal (1996), most of this discrepancy
arises from poor knowledge of molecular line opacities 
at cooler temperatures. This makes the atmosphere pseudocontinuum
at 5000-5300 \AA\ higher, and the relevant indices stronger. 
These four stars will be excluded from further analyses, and we are therefore 
left with 72 stars for the ``fair'' sample and 87 stars for the
``extended'' sample.

\begin{figure}[t]
\resizebox{\hsize}{!}{\includegraphics{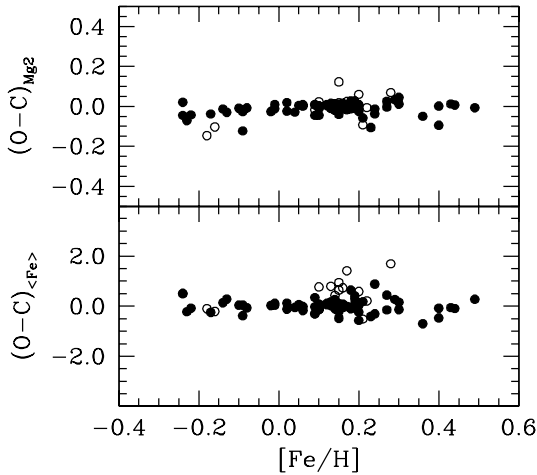}}
\caption{The metallicity scale is explored vs. (O--C) index residuals 
(in the sense [Observed -- synthetic]) like in Fig.~\ref{oc_all}.
No (O--C) trend is found vs.\ [Fe/H], with index 
residual distribution consistent with a zero average. At least 6 stars
can be confidently recognized with [Fe/H] in excess of +0.35~dex.
}
\label{feh_all}
\end{figure}

According to the fiducial atmosphere parameters of Paper I, the resulting 
metallicity scale is explored in Fig.~\ref{feh_all}. As for Fig.~\ref{oc_all},
the (O--C) distribution vs. [Fe/H] does not show any significant trend being 
consistent in any case with a zero average. At least 6 stars can be
confidently recognized on the plot with a metallicity in excess of 
[Fe/H]~$>+0.35$~dex. Some of these SMR candidates have been 
discussed in detail in Paper I. 

\subsection {Magnesium vs.\ Iron relative abundance}

A study of the Mg$_2$ and $<{\rm Fe}>$ residuals could also provide
important information about the Mg vs.\ Fe relative abundance.
As Kurucz model atmospheres and Chavez \etal (1997) synthetic spectra
assume a solar chemical partition (so that 
[Mg/Fe] = 0.0, by definition), any systematic deviation in the
index residuals could therefore track, on the average, a non-standard
chemical mix for real stars.

Two opposite trends in the (O--C) distribution might be envisaged, in this
respect, depending whether they are induced by a biased set of atmosphere 
parameters or rather by a non-solar abundance partition for [Mg/Fe].
In the first case, a mismatch of either $T_{\it eff}$ and/or $\log g$
would reflect in a nominal metallicity shift (in force of 
eqs.~(\ref{eq:e10})-(\ref{eq:e12}), and therefore in a {\it correlated}
trend of Mg$_2$ and $<{\rm Fe}>$ (O--C) residuals.
On the contrary, any {\it intrinsic} change of the [Mg/Fe] relative abundance
would induce {\it anti-correlated} (O--C)s for the two indices.

\begin{figure}[h]
\resizebox{\hsize}{!}{\includegraphics{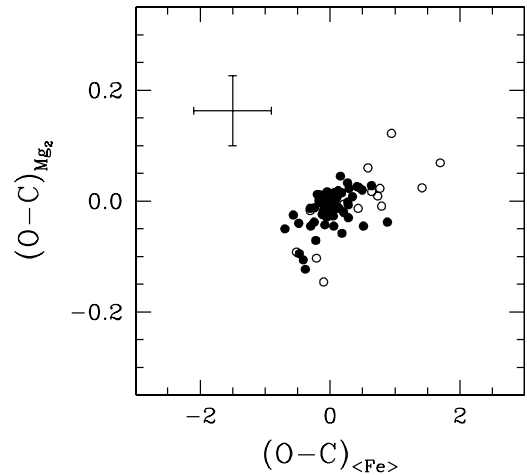}}
\caption{The Mg$_2$ and $<{\rm Fe}>$ index residuals between observations 
and fiducial synthetic spectra are displayed for the ``fair'' ($\bullet$)
and the ``extended'' ($\circ$) sample. 
The typical 2-$\sigma$ error box of individual observations is displayed
top left in the figure. The lack of any clear {\it anti-correlated}
trend between the (O--C) residuals indicates, on the average, a solar [Mg/Fe] 
relative abundance. See text for discussion.
}
\label{res_all}
\end{figure}

As far as the the ``fair'' sample is accounted for in our analysis
(cf. solid dots distribution in Fig.~\ref{res_all}) 
no statistically significant trend between Mg and Fe indices is evident
within a 2-$\sigma$ confidence level. As expected, most of the
deviating points in the figure  belong to the ``extended'' sample,
and the correlated distribution in the latter case is therefore induced 
as a result of the less accurate atmosphere fit.

In conclusion, we have therefore no evident sign of a systematic trend in 
the [Mg/Fe] relative abundance, and a standard chemical partition seems 
consistent with the data.

\section{Comparison with index fitting functions}

Lick narrow-band indices have proven to be an useful diagnostic tool not only 
to probe atmosphere physical properties in individual stars but also for
studying stellar populations as a whole (Buzzoni 1995, 1996; 
Worthey \etal 1992, 1996).

In particular, population synthesis models have made extensive use
of the ``fitting function'' technique to compute integrated indices
for theoretical stellar systems.

This procedure relies on analytical fits of the index strength vs.\ stellar 
fundamental parameters (namely $\log T_{\it eff}$, $\log g$, and [Fe/H]) 
across the H-R diagram, and allows spectroscopic features to be synthesized
for stellar aggregates even by means of low-resolution models of the
integrated SED.

For example, if an index $I$ can be defined for individual stars in a 
magnitude scale like in eq.~(\ref{eq:index}), then by summing up on the whole 
stellar population, we have
\begin{equation}
I_{tot} = -2.5 \log \left({{\sum_* f_c(i)\times 10^{-0.4~I(i)}}\over {\sum_* f_c(i)}}\right).
\end{equation}

If the dex term in the r.h.\ upper sum can be evaluated analytically for each 
composing star, then the integrated index simply follows
in terms of the total continuum luminosity  of the whole population,
$L_{tot} = \sum_* f_c(i)$ (see B92 for a first application
of this technique to the Mg$_2$ synthesis).

Analytical fitting functions for Iron and Magnesium indices have been
provided by B92, B94 and Gorgas \etal (1993). The latter set
of equations has then been revised by W94 including a larger 
number of features in the optical wavelength range and calibrating vs.
effective temperature.

It could be useful to test self-consistency of these functions 
also in the SMR regime in order to secure their application
to the synthesis of metal-rich stellar environments, like in the 
case of the Galaxy bulge or external early-type galaxies.

\begin{figure}
\resizebox{\hsize}{!}{\includegraphics{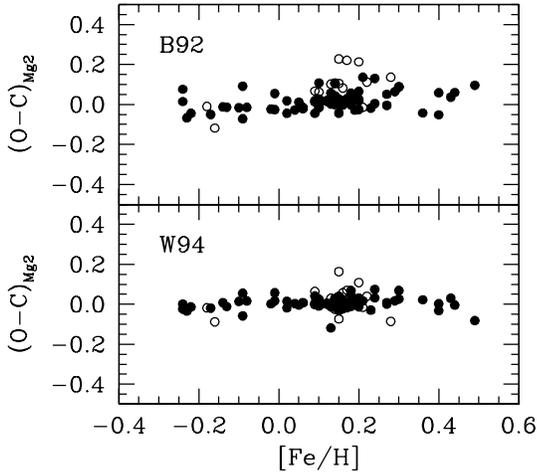}}
\caption{The Mg$_2$ index residuals (in the sense [Observed -- Computed])
with respect to B92 and W94
fitting functions vs. metallicity for the ``fair'' ($\bullet$)
and the ``extended'' ($\circ$) samples. Both (O--C) distributions are 
consistent with a zero average. Point spread for the ``fair'' sample across 
the B92 fitting function is $\sigma({\rm Mg}_2) = \pm 0.049$~mag, 
and $\sigma({\rm Mg}_2) = \pm 0.031$~mag with respect to W94.
}
\label{mg2_fit}
\end{figure}

A first check in this sense can be done by studying the index residuals
of our observations both with respect to the B92, B94 and 
W94 sets of fitting functions.
Figure~\ref{mg2_fit} summarizes our results for Mg$_2$. 

Both (O--C) distributions in the figure are consistent with a zero average 
residual confirming that the two sets of equations properly account for 
high-metallicity stars. 
The W94 fitting function is slightly more accurate than the B92 one
(point spread for the ``fair sample'' is $\sigma({\rm Mg}_2) 
= \pm 0.031$~mag vs.\ $\sigma({\rm Mg}_2) = \pm 0.049$~mag for the latter 
case), but at cost of a more elaborated analytical fit consisting of 
four different polynomials along the temperature range of 
stars.\footnote{Quite importantly, note that an incorrect version of one of the four
polynomial branches is given in the Worthey \etal (1994) Table 3.
The claimed $\log \Theta$ dependence of the fit should in fact be read
as a simple $\Theta$ dependence. The same problem affects also all the other
indices discussed in this work.}

\begin{figure}
\resizebox{\hsize}{!}{\includegraphics{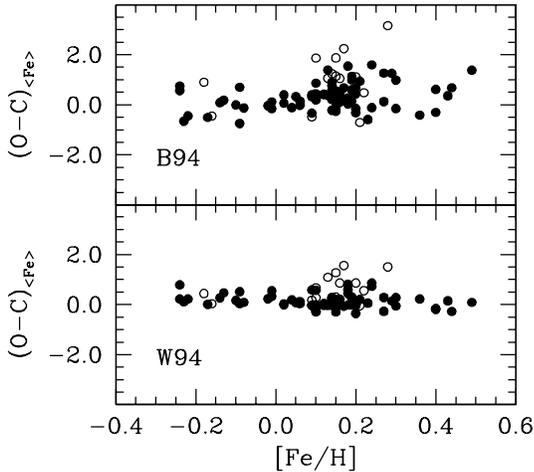}}
\caption{Same as for Fig.~\ref{mg2_fit} but for the combined Fe index
$<{\rm Fe}> = ({\rm Fe52 + Fe53})/2$.
Point spread of the ``fair sample'' ($\bullet$) with respect to the
zero-average (O--C) is $\sigma(<{\rm Fe}>) = \pm 0.61$~\AA\ comparing with
B94, and $\sigma(<{\rm Fe}>) = \pm 0.30$~\AA\ with respect to W94.
}
\label{fe_fit}
\end{figure}

A similar behaviour is also found for the $<{\rm Fe}>$ index (cf.\ Fig.~\ref{fe_fit})
with $\sigma(<{\rm Fe}>) = \pm 0.61$~\AA\ with respect to B94, and 
$\sigma(<{\rm Fe}>) = \pm 0.30$~\AA\ with respect to the W94 three-branch
fit. We verified that the skewed positive
residuals with respect to B94, evident from the figure, mostly 
come from the warmer stars in the sample ($T_{\it eff} \gtrsim 6700$~K) for which 
the fitting function predicts a vanishing Fe52 index.

\begin{figure}
\resizebox{\hsize}{!}{\includegraphics{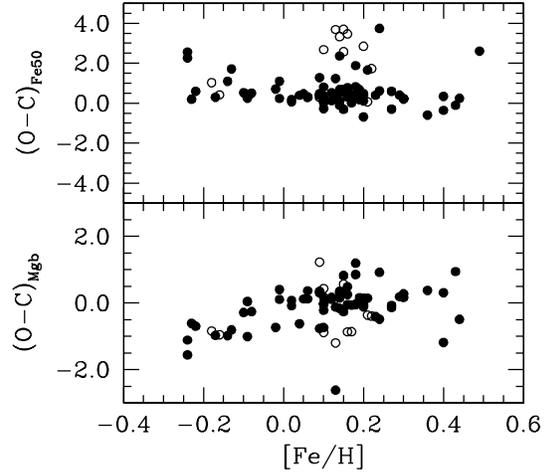}}
\caption{Same as for Fig.~\ref{mg2_fit} but for the Fe50 and Mgb indices.
The (O--C) residuals are computed with respect to the W94 fitting functions. 
Point spread for the ``fair sample'' ($\bullet$)
with respect to the zero-average (O--C) is $\sigma({\rm Fe50}) = \pm 0.95$~\AA, 
and $\sigma({\rm Mgb}) = \pm 0.83$~\AA. A trend with [Fe/H] (opposite
for Fe50 and Mgb) appears in the data. 
}
\label{fe50_fit}
\end{figure}

\begin{figure}
\resizebox{\hsize}{!}{\includegraphics{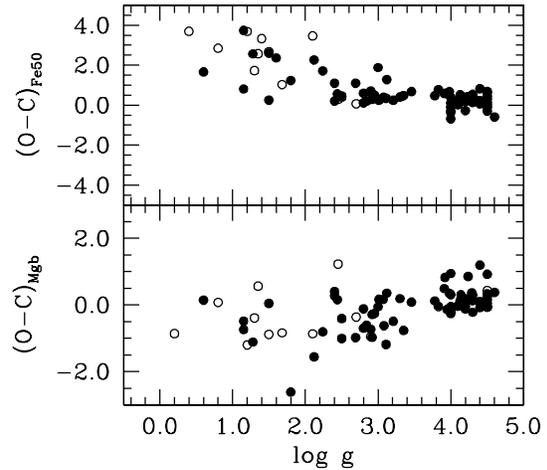}}
\caption{The Fe50 and Mgb (O--C) residuals as from Fig.~\ref{fe50_fit}
is explored vs.\ stellar gravity. The drift in the data distribution indicates
that $\log g$ dependence is not fully accounted for by the W94
fitting functions. The residual variance is the main responsible for the poorer 
match to the observations in Fig.~\ref{fe50_fit}.
}
\label{fe50g_fit}
\end{figure}

The Fe50 and Mgb features have been accounted only by W94. Fitting function
predictions are compared with observations in Fig.~\ref{fe50_fit}.
Data seem more poorly matched, compared with Mg$_2$ and $<{\rm Fe}>$,
with a more important (O--C) spread ($\sigma({\rm Fe50}) = \pm 0.95$~\AA, 
and $\sigma({\rm Mgb}) = \pm 0.83$~\AA\ for the ``fair sample''), and a 
non-zero (O--C) average.

If we explore the (O--C) distribution vs. stellar surface gravity 
(cf.\ Fig.~\ref{fe50g_fit})
the drift of the data is even increased confirming that $\log g$ dependence 
has not been fully accounted for by the W94 fitting functions.
This residual variance is expected to induce a biased measure of the integrated
Fe50 and Mgb as far as population synthesis models are computed relying
on the W94 theoretical framework. Special caution should therefore be 
recommended in the synthesis of these indices.

\subsection {Fitting functions and index fit}

Although each of the B92, B94 and W94 fitting functions assures
a convenient representation of the Mg$_2$ and $<{\rm Fe}>$ indices 
in the metallicity range of our observations, it is however worth to briefly
discuss here a possible important bias dealing with any unwarranted use of 
the fitting function technique to reproduce stellar indices.

The problem could be of paramount importance in case of population synthesis
models, where fitting functions are used to a systematic prediction of
index behaviour across the whole H-R diagram and often beyond the
boundary limits of the formal fitting domain.

\begin{figure}
\resizebox{\hsize}{!}{\includegraphics{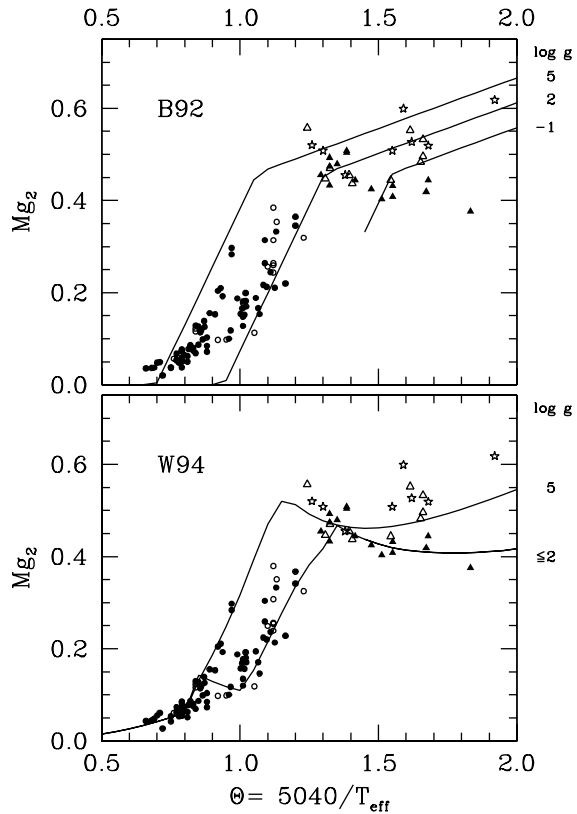}}
\caption{The Mg$_2$ data distribution ($\bullet$ = ``fair'', $\circ$ = 
``extended'' samples) vs.\ B92 and W94 fitting function predictions (solid lines).
To ease comparison, observations have been corrected to [Fe/H] = 0 
by subtracting the corresponding [Fe/H] polynomial term of each fitting set.
In order to span the whole range of temperature, also Mg$_2$ measurements of 
eight Gliese red dwarfs from B92 (``$\star$'' markers) have been added in both panels
together with ten M dwarfs from the Gorgas \etal (1993) sample
(``$\vartriangle$''). The low-gravity range is also spanned by a set of 16 field M giants 
with $T_{\it eff} < 3900$~K, from W94 (``$\blacktriangle$''). The B92 and W94 fitting 
functions have been computed for $\log g = 5.0, 2.0$ and --1.0 dex, as labelled in each 
panel. Below 3900~K the W94 fit is insensitive to stellar gravity so that the
$\log g = 2$ and --1 curves merge. A solar metallicity is assumed throughout in the
models.}
\label{ff}
\end{figure}

An instructive example in this sense is given in Fig.~\ref{ff}, where we
compare Mg$_2$ distribution of our stellar sample with 
the B92 and W94 analytical output.
In order to span the whole range of temperature, to our stars we also added the
Mg$_2$ measurements of eight Gliese red dwarfs from B92
in both panels) and ten M dwarfs from the Gorgas \etal (1993) sample
with $3000 < T_{\it eff} < 4300$~K, according to the 
$(V-K)$ vs.\ $T_{\it eff}$ calibration of W94 (cf.\ Table 6 therein).
A set of 16 field M giants with $T_{\it eff} < 3900$~K, from Table A2C of W94, 
is also included in the figure to span low-gravity stars.

To ease comparison, our observations have been corrected for metallicity 
and reconducted to [Fe/H] = 0 by subtracting the [Fe/H] polynomial term 
according to the different fitting sets.
In each panel of the figure, three theoretical loci are computed for 
$\log g = 5.0, 2.0$ and --1.0 dex at solar metallicity.

\begin{figure*}
\resizebox{\hsize}{!}{\includegraphics{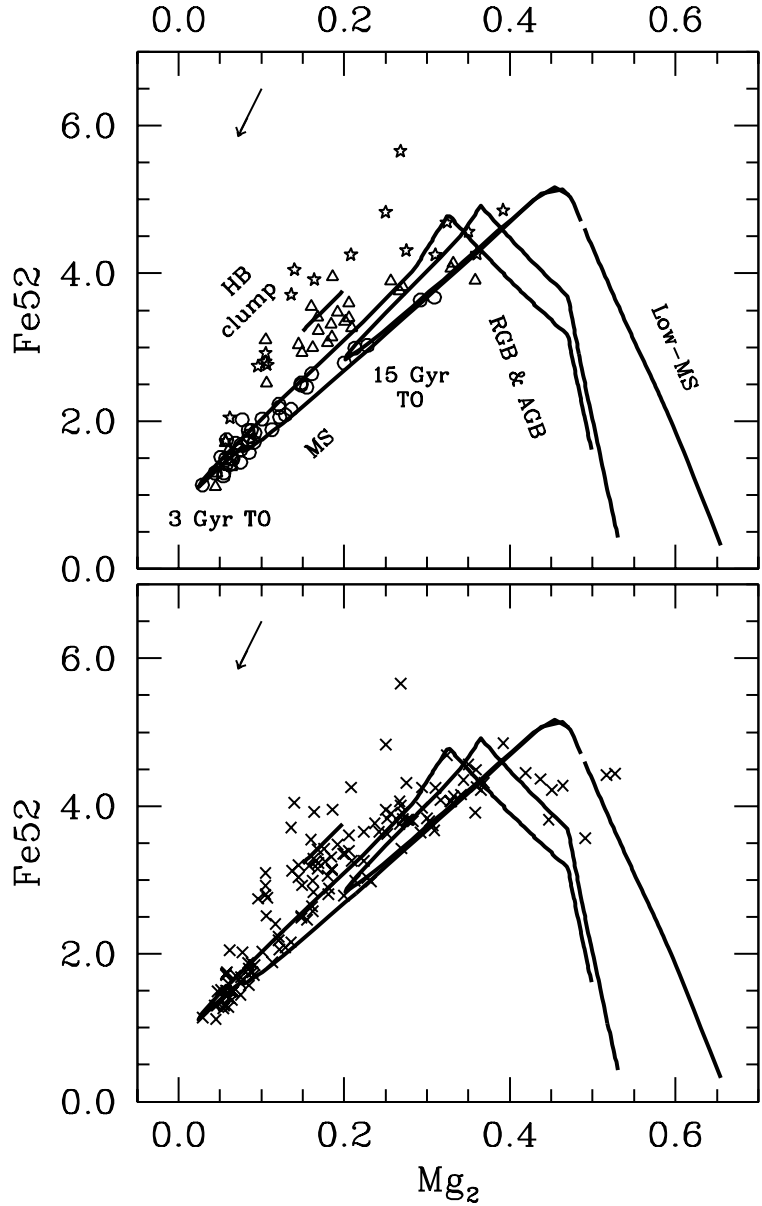}}
\caption{The Mg$_2$ vs.\ Fe52 distribution for the 87 stars with fiducial 
atmosphere parameters in the ``extended'' sample {\it (upper panel)} and 
for the global observed sample of 139 stars {\it (lower panel)}.
Stars in the upper panel are marked according to their surface gravity
($\star: \log g < 2.0$; $\vartriangle: 2.00 \leq \log g \leq 3.5$;
$\circ: \log g > 3.5$).
The theoretical locus for a 3 and 15 Gyr old simple stellar populations
with [Fe/H] = +0.2 from Buzzoni (1989) is superposed 
to the data, and main evolutionary
phases are labelled (MS = main sequence; TO = MS Turn off point; 
RGB = red giant branch; AGB = asymptotic giant branch; HB = horizontal branch).
The vector top left in each panel is the expected variation in the observed 
indices for a change in [Fe/H] of --0.5 dex.
}
\label{femg}
\end{figure*}

\clearpage
\begin{deluxetable}{rlclccccccl}
\tabletypesize{\footnotesize}
\tablewidth{0pc}
\tablenum{1}
\tablecaption{Observational database and standard Lick indices}
\tablehead{
\colhead{ HD } & \colhead{Sp. Type} & \colhead{n(obs)} & \colhead{$\sigma$(flux)} & \colhead{Fe50} & \colhead{Mg$_2$} & \colhead{Mg$_b$} & \colhead{Fe52} &\colhead{Fe53} & \colhead {[Fe/H] \tablenotemark{a}} & \colhead {Remarks \tablenotemark{b}}\\
\colhead{ } & \colhead{ } & \colhead{ } & \colhead{ } & \colhead{[\AA]} & \colhead{[mag]} & \colhead{[\AA]} & \colhead{[\AA]} &\colhead{[\AA]} & \colhead { } & \colhead { }
}
\startdata
      4 & F0          & 1 & \nodata &  4.098 & 0.075 & 1.447 & 1.682 & 1.269 & +0.3\phn  &    \\
   1461 & G0V         & 2 &  0.0048 &  4.552 & 0.148 & 3.175 & 2.487 & 1.932 & +0.43     & W  \\
   1835 & G3V         & 2 &  0.0109 &  4.722 & 0.148 & 3.019 & 2.515 & 1.968 & +0.19     & B  \\
   4188 & K0III       & 3 &  0.0068 &  5.952 & 0.105 & 2.548 & 3.096 & 2.491 & --0.16    &    \\
   6497 & K2III       & 3 &  0.0061 &  6.294 & 0.307 & 4.131 & 3.736 & 3.351 &  \nodata  &    \\
   8673 & F7V         & 2 &  0.0037 &  3.754 & 0.085 & 1.505 & 1.574 & 1.233 & +0.16     &    \\
  10307 & G1.5V       & 3 &  0.0068 &  3.766 & 0.136 & 2.610 & 2.160 & 1.565 & --0.02    & W  \\
  10780 & K0V         & 2 &  0.0046 &  4.690 & 0.228 & 4.677 & 3.026 & 2.441 & +0.36     & W  \\
  15152 & K5III       & 2 &  0.0095 &  7.602 & 0.437 & 4.389 & 4.366 & 4.385 &  \nodata  &    \\
  16232 & F6V         & 2 &  0.0051 &  3.284 & 0.062 & 1.119 & 1.408 & 0.984 & +0.27     &    \\
        &             &   &         &        &       &       &       &       &           &    \\
  17017 & K2III       & 2 &  0.0076 &  6.399 & 0.252 & 3.513 & 3.621 & 3.149 &  \nodata  &    \\
  18322 & K1III       & 2 &  0.0056 &  5.838 & 0.201 & 3.197 & 3.353 & 2.788 & --0.23    & B  \\
  19476 & K0III       & 4 &  0.0068 &  6.080 & 0.184 & 2.826 & 3.312 & 2.749 & +0.04     & W  \\
  20630 & G5V         & 2 &  0.0080 &  4.245 & 0.155 & 3.049 & 2.460 & 1.912 & --0.01    & W B\\
  20675 & F6V         & 3 &  0.0047 &  3.725 & 0.065 & 1.079 & 1.485 & 1.170 & +0.2\phn  &    \\
  21488 & K0          & 4 &  0.0159 &  5.023 & 0.162 & 2.164 & 2.574 & 2.276 &  \nodata  &    \\
  24240 & K0III       & 3 &  0.0065 &  6.730 & 0.176 & 2.597 & 3.423 & 2.894 &  \nodata  &    \\
  24802 & K0          & 3 &  0.0081 &  7.159 & 0.250 & 2.782 & 3.949 & 3.396 &  \nodata  &    \\
  25975 & K1III       & 3 &  0.0052 &  5.416 & 0.224 & 3.772 & 3.261 & 2.681 &  \nodata  &    \\
  26462 & F4V         & 3 &  0.0016 &  3.142 & 0.060 & 0.901 & 1.278 & 0.965 &  \nodata  & W  \\
        &             &   &         &        &       &       &       &       &           &    \\
  26846 & K3III       & 2 &  0.0041 &  6.741 & 0.267 & 3.843 & 3.758 & 3.352 & +0.21     &    \\
  27371 & K0III       & 3 &  0.0060 &  6.333 & 0.169 & 2.459 & 3.228 & 2.703 & --0.02    & W  \\
  27697 & K0I         & 3 &  0.0017 &  6.349 & 0.177 & 2.434 & 3.251 & 2.699 &  \nodata  & W  \\
  28307 & K0IIIb      & 3 &  0.0019 &  6.124 & 0.185 & 2.637 & 3.148 & 2.654 &  \nodata  & W  \\
  30495 & G3V         & 2 &  0.0026 &  3.932 & 0.121 & 2.857 & 2.181 & 1.676 & +0.1\phn  & B  \\
  30562 & F8V         & 2 &  0.0032 &  4.708 & 0.121 & 2.390 & 2.229 & 1.774 & +0.14     & B  \\
  30652 & F6V         & 3 &  0.0039 &  3.341 & 0.075 & 1.326 & 1.445 & 1.101 & +0.02     & W  \\
  32068 & K4Ib-II+... & 2 &  0.0101 &  8.411 & 0.350 & 3.018 & 4.563 & 4.453 & +0.1\phn  &    \\
  32393 & K3          & 1 & \nodata &  7.057 & 0.328 & 4.093 & 4.063 & 3.702 &  \nodata  &    \\
  33276 & F2IV        & 2 &  0.0076 &  3.342 & 0.064 & 0.989 & 1.381 & 1.116 & +0.29     &    \\
        &             &   &         &        &       &       &       &       &           &    \\
  34411 & G1.5IV-V    & 2 &  0.0034 &  4.212 & 0.122 & 2.523 & 2.057 & 1.661 & +0.06     & W  \\
  35620 & K3IIICN+... & 1 & \nodata &  7.232 & 0.360 & 3.721 & 4.262 & 4.046 & --0.09    & W  \\
  36040 & K0IIIp      & 1 & \nodata &  6.828 & 0.242 & 3.321 & 3.650 & 3.134 &  \nodata  &    \\
  36389 & M2Iab       & 2 &  0.0066 &  7.183 & 0.232 & 2.889 & 2.977 & 3.005 & +0.11     &    \\
  37387 & K1Ib        & 1 & \nodata & 10.895 & 0.250 & 1.484 & 4.831 & 4.433 & +0.13     &    \\
  37763 & K3III       & 1 & \nodata &  7.135 & 0.329 & 4.119 & 4.068 & 3.742 & +0.3\phn  & B  \\
  42341 & K2III       & 1 & \nodata &  7.144 & 0.294 & 4.154 & 3.980 & 3.670 &  \nodata  &    \\
  44391 & K0Ib        & 2 &  0.0111 &  8.916 & 0.164 & 1.373 & 3.919 & 3.278 & +0.21     &    \\
  44478 & M3II        & 2 &  0.0095 & 10.685 & 0.491 & 7.131 & 3.563 & 3.502 &  \nodata  &    \\
  45412 & F8Ib var    & 2 &  0.0046 &  5.721 & 0.062 & 0.187 & 2.048 & 1.682 & +0.1\phn  &    \\
        &             &   &         &        &       &       &       &       &           &    \\
  46709 & K4III       & 2 &  0.0101 &  7.900 & 0.359 & 3.444 & 4.492 & 4.297 &  \nodata  &    \\
  47174 & K3Iab:      & 2 &  0.0046 &  6.988 & 0.237 & 2.958 & 3.766 & 3.226 &  \nodata  &    \\
  48682 & G0V         & 2 &  0.0083 &  3.972 & 0.092 & 2.106 & 1.845 & 1.475 & +0.15     & W  \\
  49161 & K4III       & 2 &  0.0052 &  7.931 & 0.344 & 3.709 & 4.353 & 4.248 &  \nodata  & W  \\
  52973 & G0Ib var    & 2 &  0.0078 &  7.722 & 0.096 & 0.258 & 2.745 & 2.083 & +0.49     &    \\
  54719 & K2III       & 2 &  0.0071 &  7.511 & 0.267 & 3.405 & 4.065 & 3.577 &  \nodata  & W  \\
  56577 & K4III       & 1 & \nodata &  9.760 & 0.392 & 3.437 & 4.853 & 4.814 & +0.15     & B  \\
  57727 & G8III       & 2 &  0.0045 &  5.175 & 0.163 & 2.536 & 2.833 & 2.189 &  \nodata  &    \\
  58207 & G9IIIb      & 2 &  0.0064 &  5.832 & 0.180 & 2.600 & 3.057 & 2.534 & --0.17    & W  \\
  59881 & F0III       & 3 &  0.0077 &  2.642 & 0.045 & 0.727 & 1.115 & 0.864 & +0.19     &    \\
  60522 & M0I         & 2 &  0.0331 &  5.268 & 0.451 & 5.010 & 4.220 & 2.926 &  \nodata  & W  \\
  61064 & F6III       & 2 &  0.0107 &  4.308 & 0.060 & 0.761 & 1.602 & 1.259 & +0.44     &    \\
  62902 & K5III       & 2 &  0.0091 &  7.668 & 0.419 & 4.766 & 4.445 & 4.531 &  \nodata  &    \\
  63302 & K3Iab/Ib    & 1 & \nodata & 13.493 & 0.268 & 0.804 & 5.656 & 4.952 & +0.17     & B  \\
  63700 & G3Ib        & 1 & \nodata & 10.381 & 0.140 & 0.341 & 4.049 & 3.527 & +0.24     &    \\
  65714 & gG8         & 2 &  0.0100 &  6.773 & 0.163 & 2.299 & 3.288 & 2.735 &  \nodata  &    \\
  72184 & K2III       & 2 &  0.0042 &  6.613 & 0.278 & 4.028 & 3.813 & 3.322 &  \nodata  & W  \\
  72324 & G9I         & 2 &  0.0070 &  6.246 & 0.168 & 2.348 & 3.150 & 2.544 &  \nodata  & W  \\
  72505 & K0III       & 2 &  0.0085 &  6.849 & 0.300 & 3.852 & 3.837 & 3.455 &  \nodata  &    \\
  72561 & G5III       & 2 &  0.0068 &  6.801 & 0.145 & 1.573 & 3.203 & 2.583 &  \nodata  &    \\
        &             &   &         &        &       &       &       &       &           &    \\
  73665 & K0I         & 2 &  0.0071 &  6.672 & 0.166 & 2.382 & 3.256 & 2.685 &  \nodata  & W  \\
  73710 & K0I         & 2 &  0.0303 &  4.822 & 0.117 & 1.399 & 2.407 & 1.876 &  \nodata  & W  \\
  74739 & G7.5IIIa    & 2 &  0.0117 &  6.479 & 0.145 & 1.722 & 3.038 & 2.454 & --0.14    &    \\
  75732 & G8V         & 3 &  0.0059 &  5.980 & 0.309 & 5.522 & 3.674 & 3.089 & +0.24     & W  \\
  78249 & K1IV        & 2 &  0.0121 &  5.558 & 0.269 & 4.354 & 3.427 & 2.841 &  \nodata  &    \\
  81029 & F0          & 1 & \nodata &  3.298 & 0.051 & 0.565 & 1.515 & 1.015 & +0.27     &    \\
  81873 & K0III       & 2 &  0.0068 &  6.245 & 0.199 & 2.893 & 3.357 & 2.806 &  \nodata  &    \\
  82734 & K0IV        & 1 & \nodata &  6.718 & 0.168 & 2.412 & 3.411 & 2.908 & +0.4\phn  &    \\
  83951 & F3V         & 2 &  0.0045 &  3.130 & 0.044 & 0.853 & 1.299 & 0.975 & +0.14     &    \\
  85503 & K0III       & 2 &  0.0080 &  7.532 & 0.332 & 4.286 & 4.135 & 3.817 & --0.01    & W  \\
        &             &   &         &        &       &       &       &       &           &    \\
  87822 & F4V         & 2 &  0.0087 &  3.489 & 0.063 & 1.210 & 1.491 & 1.165 & +0.19     &    \\
  88230 & K8V         & 2 &  0.0137 &  5.126 & 0.527 & 4.759 & 4.442 & 4.106 & +0.28     & W  \\
  88284 & K0III       & 1 & \nodata &  6.452 & 0.192 & 3.049 & 3.478 & 2.963 & +0.09     & W B\\
  90277 & F0V         & 2 &  0.0082 &  3.129 & 0.046 & 0.739 & 1.297 & 0.960 & +0.19     &    \\
  92125 & G2.5IIa     & 2 &  0.0129 &  6.431 & 0.106 & 1.088 & 2.515 & 2.168 & --0.24    &    \\
  93257 & gK3         & 2 &  0.0139 &  6.408 & 0.303 & 4.284 & 3.796 & 3.337 &  \nodata  &    \\
  95272 & K0III       & 1 & \nodata &  6.216 & 0.206 & 2.965 & 3.397 & 2.818 & --0.22    & W B\\
  95849 & K3III       & 2 &  0.0106 &  7.243 & 0.270 & 3.560 & 3.909 & 3.559 &  \nodata  &    \\
 100563 & F5V         & 2 &  0.0087 &  3.488 & 0.066 & 1.418 & 1.489 & 1.203 & +0.12     &    \\
 101013 & K0IIIp      & 1 & \nodata &  5.099 & 0.181 & 1.888 & 2.809 & 1.980 &  \nodata  &    \\
        &             &   &         &        &       &       &       &       &           &    \\
 102328 & K3III       & 2 &  0.0171 &  7.468 & 0.365 & 4.487 & 4.216 & 4.019 &  \nodata  & W  \\
 102634 & F7V         & 2 &  0.0055 &  4.060 & 0.085 & 1.782 & 1.798 & 1.373 & +0.1\phn  & B  \\
 102870 & F9V         & 2 &  0.0044 &  4.127 & 0.084 & 1.814 & 1.876 & 1.429 & +0.18     & W B\\
 104304 & G9IV        & 2 &  0.0045 &  5.174 & 0.213 & 4.120 & 2.989 & 2.400 & +0.18     &    \\
 109511 & K2III       & 2 &  0.0089 &  6.663 & 0.206 & 2.834 & 3.603 & 3.057 & --0.09    &    \\
 110014 & K2III       & 1 & \nodata &  7.518 & 0.294 & 3.617 & 4.252 & 3.769 &  \nodata  &    \\
 113022 & F6Vs        & 1 & \nodata &  3.576 & 0.057 & 1.432 & 1.500 & 1.109 & +0.1\phn  &    \\
 114710 & F9.5V       & 1 & \nodata &  3.775 & 0.088 & 2.282 & 1.880 & 1.450 & +0.02     & W  \\
 115604 & F3III       & 1 & \nodata &  4.506 & 0.057 & 0.693 & 1.713 & 1.373 & +0.18     &    \\
 120136 & F6IV        & 1 & \nodata &  3.940 & 0.058 & 1.544 & 1.750 & 1.326 & +0.14     & W  \\
        &             &   &         &        &       &       &       &       &           &    \\
 121370 & G0IV        & 1 & \nodata &  4.704 & 0.077 & 1.700 & 2.022 & 1.585 & +0.16     & W  \\
 124425 & F7IV        & 1 & \nodata &  3.587 & 0.047 & 1.010 & 1.496 & 1.044 &  \nodata  &    \\
 124570 & F6IV        & 1 & \nodata &  3.952 & 0.069 & 1.517 & 1.702 & 1.325 & +0.12     &    \\
 125560 & K3III       & 1 & \nodata &  7.148 & 0.316 & 4.028 & 4.080 & 3.695 &  \nodata  & W  \\
 127227 & K5III       & 1 & \nodata &  7.608 & 0.464 & 4.680 & 4.275 & 4.491 &  \nodata  &    \\
 129989 & K0II-III    & 1 & \nodata &  7.623 & 0.160 & 2.054 & 3.549 & 3.004 & --0.13    &    \\
 130948 & G1V         & 1 & \nodata &  3.712 & 0.113 & 2.327 & 1.886 & 1.413 & +0.2\phn  &    \\
 136028 & K5I         & 1 & \nodata &  4.126 & 0.341 & 4.093 & 4.156 & 2.102 &  \nodata  & W  \\
 139357 & gK4         & 1 & \nodata &  7.535 & 0.268 & 3.807 & 4.010 & 3.709 &  \nodata  &    \\
 140573 & K2IIIb      & 1 & \nodata &  7.236 & 0.256 & 3.693 & 3.898 & 3.548 & +0.23     & W  \\
 144284 & F8IV        & 4 &  0.0063 &  4.098 & 0.075 & 1.447 & 1.682 & 1.269 & +0.2\phn  &    \\
 145000 & K1III       & 1 & \nodata &  6.841 & 0.224 & 3.366 & 3.649 & 3.219 &  \nodata  &    \\
 145675 & K0V         & 2 &  0.0101 &  6.036 & 0.292 & 5.508 & 3.640 & 3.158 & +0.18     & W  \\
 146051 & M0.5III     & 2 &  0.0093 &  7.362 & 0.447 & 4.757 & 3.818 & 3.970 & +0.32     &    \\
 147677 & K0III       & 1 & \nodata &  6.039 & 0.158 & 2.633 & 3.195 & 2.651 &  \nodata  & W  \\
 148513 & K4III       & 1 & \nodata &  7.460 & 0.369 & 3.948 & 4.312 & 4.276 &  \nodata  & W  \\
 150680 & G0IV        & 2 &  0.0055 &  4.390 & 0.101 & 2.022 & 2.030 & 1.636 & --0.07    &    \\
 153956 & gK1         & 1 & \nodata &  6.821 & 0.276 & 3.564 & 3.812 & 3.324 &  \nodata  &    \\
 156266 & K2III       & 1 & \nodata &  6.692 & 0.252 & 3.512 & 3.830 & 3.320 &  \nodata  &    \\
 156283 & K3Iab:      & 2 &  0.0100 &  7.798 & 0.310 & 3.083 & 4.250 & 3.899 & --0.18    &    \\
        &             &   &         &        &       &       &       &       &  \nodata  &    \\
 157881 & K7V         & 2 &  0.0067 &  5.017 & 0.517 & 4.687 & 4.419 & 4.156 & +0.4\phn  &    \\
 159181 & G2Iab:      & 2 &  0.0075 &  7.377 & 0.107 & 0.735 & 2.763 & 2.394 & +0.14     &    \\
 159925 & G9III       & 1 & \nodata &  6.117 & 0.137 & 2.129 & 3.123 & 2.432 &  \nodata  &    \\
 160922 & F5V         & 2 &  0.0091 &  3.211 & 0.058 & 1.323 & 1.423 & 1.054 & +0.4\phn  &    \\
 161096 & K2III       & 2 &  0.0146 &  7.130 & 0.271 & 3.866 & 3.824 & 3.494 & +0.14     &    \\
 161797 & G5IV        & 3 &  0.0053 &  5.166 & 0.161 & 3.128 & 2.641 & 2.112 & +0.16     & W  \\
 162917 & F4IV-V      & 4 &  0.0049 &  3.343 & 0.055 & 1.241 & 1.302 & 0.865 & +0.1\phn  &    \\
 163770 & K1IIaCN+... & 1 & \nodata &  9.177 & 0.208 & 1.748 & 4.253 & 3.827 & --0.24    &    \\
 163993 & G8III       & 3 &  0.0056 &  5.796 & 0.149 & 2.498 & 2.928 & 2.342 & --0.1\phn &    \\
 166229 & K2.5III     & 1 & \nodata &  6.535 & 0.283 & 3.936 & 3.798 & 3.386 &  \nodata  &    \\
        &             &   &         &        &       &       &       &       &           &    \\
 167858 & F2V         & 3 &  0.0067 &  2.805 & 0.029 & 0.792 & 1.137 & 0.826 & +0.17     &    \\
 171802 & F5III       & 2 &  0.0032 &  2.949 & 0.054 & 1.151 & 1.259 & 0.772 & +0.1\phn  &    \\
 181276 & G9III       & 2 &  0.0055 &  5.863 & 0.162 & 2.696 & 2.990 & 2.284 & --0.08    &    \\
 182572 & G8IV...     & 2 &  0.0040 &  5.408 & 0.200 & 3.788 & 2.787 & 2.271 & +0.15     & W  \\
 186408 & G1V         & 2 &  0.0043 &  4.281 & 0.129 & 2.931 & 2.085 & 1.604 & +0.06     & W  \\
 187238 & K3Ia0-Ia    & 3 &  0.0054 & 10.377 & 0.324 & 2.424 & 4.690 & 4.407 & +0.2\phn  &    \\
 187299 & G5Ia0-Ib    & 2 &  0.0054 &  9.634 & 0.186 & 1.179 & 3.954 & 3.374 & +0.16     &    \\
 187691 & F8V         & 2 &  0.0037 &  3.958 & 0.091 & 1.987 & 1.714 & 1.285 & +0.09     & W  \\
 187921 & G2.5:Iab    & 3 &  0.0101 & 10.689 & 0.136 &-0.425 & 3.711 & 2.975 & +0.28     &    \\
 196725 & K3Iab       & 2 &  0.0070 &  9.088 & 0.275 & 2.517 & 4.311 & 4.142 & +0.22     &    \\
        &             &   &         &        &       &       &       &       &           &    \\
 197039 & F5          & 2 &  0.0034 &  3.926 & 0.075 & 1.465 & 1.607 & 1.164 & +0.15     &    \\
 197572 & F7Ib...     & 3 &  0.0057 &  8.352 & 0.105 & 0.417 & 2.804 & 2.215 & +0.15     &    \\
 197963 & F7V         & 1 & \nodata &  5.464 & 0.182 & 3.030 & 2.886 & 2.516 &  \nodata  &    \\
 198084 & F8IV-V      & 2 &  0.0032 &  4.106 & 0.084 & 1.739 & 1.727 & 1.388 & +0.12     &    \\
 201078 & F7.5Ib-IIvar& 2 &  0.0039 &  4.803 & 0.057 & 0.406 & 1.735 & 1.203 & +0.13     &    \\
 205512 & K0.5III     & 2 &  0.0050 &  6.232 & 0.209 & 3.091 & 3.262 & 2.793 & +0.2\phn  &    \\
 209750 & G2Ib        & 2 &  0.0049 &  8.247 & 0.105 & 0.531 & 2.920 & 2.462 & +0.14     &    \\
 216228 & K0III       & 2 &  0.0041 &  6.104 & 0.185 & 2.908 & 3.136 & 2.626 & +0.09     &    \\
 221148 & K3IIIvar    & 2 &  0.0059 &  6.925 & 0.358 & 5.316 & 3.909 & 3.628 & +0.09     & W B\\
\enddata
\tablenotetext{a} {From Paper I}
\tablenotetext{b}{Index calibrator in common with Worthey  et al. (1994) [W] or Buzzoni et al. (1992, 1994) [B]}
\end{deluxetable}

\clearpage

While both fitting sets properly comprise our data distribution
(thus confirming the adequacy of the fit, as we discussed
in previous section), it is however evident that a different
trend for the Mg$_2$ index is predicted at low temperature.

Compared with W94, for $\Theta \gtrsim 1.3$ ($T_{\it eff} \lesssim 3900$~K) 
the B92 fitting function shows a steeper increase in the Mg index.
This better matches the red-dwarf data, although some marginal evidence
exists for an overestimate of the M giants.
On the other hand, the W94 fit predicts slightly lower values for Mg$_2$ 
throughout in the low-temperature range, and partially misses the red-dwarf
data. Moreover, an artificial glitch is evident in one of the fitting branches of
the W94 set at about $\Theta = 0.8 - 1.0$ for $\log g = 2$.

Such a different behaviour in the H-R diagram description directly reflects
in the theoretical output when computing integrated indices for
population synthesis models. This explain, for instance, the systematically
lower values for Mg$_2$ in the Worthey (1994) simple stellar population
models compared with B92.

\section{The Mg$_2$ vs.\ Fe52 diagnostics}

Given a selective dependence on the atmosphere fundamental parameters,
a combined study of the Mg and Fe Lick indices could give direct hints
on the distinctive parameters of stars.

A two-index plot, like in Fig.~\ref{femg}, summarizes in facts most of the
relevant features of a c-m diagram but with a major advantage to be both
reddening free and distance independent, and points therefore to the intrinsic
properties of stars.

In the upper panel of the figure we study the ``fair sample'' distribution 
in the Mg$_2$ vs.\ Fe52 index domain with two [Fe/H] = +0.22~dex isochrones 
of 3 and 15 Gyr from the Buzzoni (1989) synthesis code. Surface gravity of 
stars in our sample has been singled out for a better comparison with the 
models, while the main evolutionary phases for both isochrones have been 
labelled on the plot.

A quite good agreement exists between high-gravity stars ($\log g > 3.5$~dex)
and the main sequence locus confirming that objects as young as 3 Gyr are 
present in our sample.
This makes a Mg$_2$/Fe52 plot a simple and very powerful tool to 
estimate age of stellar populations. 

Stars of intermediate gravity
($2.0 \geq \log g \geq 3.5$) consistently ``bunch'' around the expected
locus for the core Helium-burning stars (both the horizontal branch phase
of low-mass stars or the first blue loop of high-mass objects) while
a lack of objects close to the tip of the red- and asymptotic giant 
branches (that is for a vanishing value of Fe52 with Mg$_2 \gtrsim 0.4$ in 
the plot) seem to indicate a prevailing presence in our sample of high-mass 
stars ($M > 2 M_\odot$) developing a non-He-degenerate red giant phase.

\acknowledgments
It is a pleasure to thank Guy Worthey, the referee of this paper, for his
valuable input and important suggestions.
This project received partial financial support from the Italian MURST
under COFIN'98 02-013, COFIN'00 02-016 and 60\% grants, and from the 
Mexican CONACyT via grant 28506-E.


\begin{thebibliography}{}

\bibitem[]{}Brodie, J. P., and Huchra, J. P. 1990, \apj, 362, 503
\bibitem[]{}Buzzoni, A. 1989, \apjs, 71, 817
\bibitem[]{}Buzzoni, A. 1995, \apjs, 98, 69
\bibitem[]{}Buzzoni, A. 1996 in Fresh Views of Elliptical Galaxies, ed. A. Buzzoni, A. Renzini and A. Serrano (San Francisco: ASP) p. 189
\bibitem[]{}Buzzoni, A., Gariboldi, G., and  Mantegazza, L. 1992, \aj, 103, 1814 (B92)
\bibitem[]{}Buzzoni, A., Mantegazza, L., and Gariboldi, G. 1994, \aj, 107, 513 (B94)
\bibitem[]{}Cayrel de Strobel, G., Soubiran, C., Friel, E. D., Ralite, N., and Fran\c cois, P. 1997, \aaps, 124, 299
\bibitem[]{}Chavez, M., Malagnini, M. L., and Morossi, C. 1996, \apj, 471, 726
\bibitem[]{}Chavez, M., Malagnini, M. L., and Morossi, C. 1997, \aaps, 126, 267
\bibitem[]{}Faber, S. M., Burstein, D., and Dressler, A. 1977, \aj, 82, 941
\bibitem[]{}Faber, S. M., Friel, E. D., Burstein, D., and Gaskell, C. M. 1985, \apjs, 57, 711
\bibitem[]{}Frogel, J. A. 1999, \apss, 265, 303
\bibitem[]{}Gorgas, J., Faber, S. M., Burstein, D., Gonzalez, J. J., Courteau, S., and Prosser, C. 1993, \apjs, 86, 153
\bibitem[]{}Kurucz, R. L. 1993, CD-ROM 13, ATLAS9 Stellar Atmosphere Programs and 2 km/s Grid (Cambridge: SAO)
\bibitem[]{}Malagnini, M. L., Morossi, C., Buzzoni, A., and Chavez, M. 2000, \pasp, 112, 1455 (Paper I)
\bibitem[]{}McWilliam, A. 1997, \araa, 35, 503
\bibitem[]{}Smith, G., and Ruck, M. J. 2000, \aap, 356, 570
\bibitem[]{}Taylor, B. J. 1991, \apjs, 76, 715
\bibitem[]{}Taylor, B. J. 1999, \aap, 344, 655
\bibitem[]{}Worthey, G. 1994, \apjs, 95, 107
\bibitem[]{}Worthey, G., Faber, S. M., and Gonzalez, J. J. 1992, \apj, 398, 69
\bibitem[]{}Worthey, G., Faber, S. M. , Gonzalez, J. J. , and Burstein, D. 1994, \apjs, 94, 687 (W94)
\bibitem[]{}Worthey, G., Trager, S. C., and Faber, S. M. 1996 in Fresh Views of Elliptical Galaxies, ed. A. Buzzoni, A. Renzini and A. Serrano (San Francisco: ASP) p. 203
\end{thebibliography}
\end{document}